\documentclass[prd,amsfonts,nofootinbib,twocolumn,nofootinbib,amsmath,amssymb]{revtex4-1}

\usepackage[T1]{fontenc} 
\usepackage{color}
\usepackage{float} 
\usepackage{physics}
\usepackage{graphicx}
\usepackage{amsthm}
\usepackage{mathtools}
\usepackage{hyperref}
\usepackage{bm}
\usepackage{xcolor}

\hypersetup{
    colorlinks=true,
    linkcolor=red,
    citecolor=blue,
    urlcolor=black
}

\definecolor{nicegreen}{rgb}{0.1,0.5,0.1}
\definecolor{darkred}{rgb}{0.6,0.1,0.1}
\definecolor{coral}{rgb}{1.0,0.5,0.313}

\def\be{\begin{equation}}
\def\ee{\end{equation}}
\newcommand{\bea}{\begin{eqnarray}}
\newcommand{\eea}{\end{eqnarray}}
\def\bfig{\begin{figure}}
\def\efig{\end{figure}}

\def\dd{\text{d}}

\def\g{\gamma}
\def\d{\delta}
\def\l{\lambda}
\def\s{\sigma}

\def\o{\omega}

\def\k{\kappa}
\def\z{\zeta}
\def\tresid{\mathbf{\delta t}}

\def\sinc{\text{sinc}}

\def\TC{\textsc{TrueCovariance}}
\def\DP{\textsc{DiagonalPhi}}
\def\SP{\textsc{SincPhi}}
\def\FFT{\textsc{FFTInt}}

\begin{document}

\title{Beyond diagonal approximations: improved covariance modeling \\ for pulsar timing array data analysis}

\author{Marco Crisostomi$^{1,2}$}
\email{marco.crisostomi@unipi.it}
\author{Rutger van Haasteren$^{3}$}
\email{rutger@vhaasteren.com}
\author{Patrick M. Meyers$^{4,2}$}
\email{pmeyers@phys.ethz.ch}
\author{Michele Vallisneri$^{4,2}$}
\email{mvallisneri@ethz.ch}
\affiliation{${}^{1}$Dipartimento di Fisica, Universit\`a di Pisa, Largo B. Pontecorvo 3, 56127 Pisa, Italy}
\affiliation{${}^{2}$TAPIR, California Institute of Technology, Pasadena, CA 91125, USA}
\affiliation{${}^{3}$Max-Planck-Institut f{\"u}r Gravitationsphysik (Albert-Einstein-Institut), Callinstra{\ss}e 38, D-30167, Hannover, Germany\\
    Leibniz Universit{\"a}t Hannover, D-30167, Hannover, Germany}
\affiliation{${}^{4}$Department of Physics, ETH Z{\"u}rich, Wolfgang‐Pauli‐Strasse 27, 8093 Z{\"u}rich, Switzerland}

\begin{abstract}
Pulsar Timing Array (PTA) searches for nHz gravitational-wave backgrounds (GWBs) typically model time-correlated noise by assuming a diagonal covariance in Fourier space, neglecting inter-frequency correlations introduced by the finite observation window. We show that this diagonal approximation can produce method-dependent shifts in estimates of spectral parameters, especially for pulsars with shallow intrinsic red noise spectra. To address these limitations, we present a method that (i) computes the time-domain autocorrelation on a coarse grid using a fast Fourier transform (FFT), (ii) interpolates it accurately to the unevenly sampled observation times, and (iii) incorporates it into a low-rank likelihood via the Sherman--Morrison--Woodbury identity. Using both analytic covariance comparisons and end-to-end simulations inspired by the NANOGrav 15-year dataset, we demonstrate that our method captures frequency correlations faithfully, avoids Gibbs ringing, and yields estimates closer to the injected values in the realization studied with modest computational cost. As PTA datasets increase in sensitivity and complexity, our approach offers a practical and scalable path to fully accurate covariance modeling for current and future analyses.
\end{abstract}

\maketitle

\section{Introduction}
\label{sec:intro}

Pulsar Timing Arrays (PTAs) provide a unique means to probe gravitational waves (GWs) through the long-term monitoring of the times of arrival (TOAs) of radio pulses from millisecond pulsars \cite{romani1989timing,fb90}. These TOAs
can be searched for the imprint of a stochastic gravitational-wave background (GWB) \cite{hd83} produced by, for example, an ensemble of astrophysical binary black holes \cite{rsg2015}, or by relic processes from the early Universe. Given that the GWB effect is weak and easily confused with low-frequency noise, robust detection relies on the cross-correlation of signals from multiple pulsars \citep[][]{2009PhRvD..79h4030A,2015PhRvD..91d4048C,2018PhRvD..98d4003V}, modeled under the assumption that the underlying stochastic process is Gaussian and statistically stationary~\cite{Romano:2016dpx, taylor2021nanohertz}.

A key element in PTA data analysis is the efficient and accurate modeling of the covariance matrix of the timing residuals. In practice, several complicating factors arise that render the task nontrivial: (i) the power spectra of the relevant signals are extremely steep (often following a power law) and are heavily dominated by low-frequency power; (ii) the sampling of the pulsar observations is highly irregular; and (iii) the timing residuals have been pre-processed by projecting out deterministic timing-model components via a least-squares fit. To overcome the computational challenges inherent in directly inverting the full covariance matrix over many thousands of unevenly sampled points, PTA analyses typically employ a low-rank representation of the low-frequency model components, in which covariance is expanded in a Fourier basis. In the standard formulation the Fourier-space covariance is taken to be diagonal (i.e., no frequency correlations are included), an approximation that is widely used and often adequate \cite{vanHaasterenVallisneri2015,Hazboun:2019vhv,Depta:2024ykq, Pitrou:2024scp,Bernardo:2024tde} but that may not be accurate enough and could produce systematic shifts in the inference of spectral parameters.

In this work we develop an improved method to model the full frequency-correlated covariance matrix. Our approach combines the efficiency of a rank-reduced representation with the fidelity of direct time-domain methods, by employing the fast Fourier transform (FFT) and a localized interpolation scheme. Our proposed method not only accounts for the intrinsic correlations induced by finite-window effects (which are closely related to the approximation of Fourier integrals as Fourier series), but also mitigates Gibbs ringing artifacts inherent in any Fourier-based approach when severely reducing the number of frequency modes. In Sec.~\ref{sec:cov-matrix} we review the traditional Fourier-sum approach and then introduce our new method. In Sec.~\ref{sec:comparison} we compare the performance of various covariance approximations. In Sec.~\ref{PTA} we apply our method to simulated PTA datasets inspired by the NANOGrav 15-year data release. In Sec.~\ref{sec:conclusion} we offer our conclusions.

We have implemented the methods discussed in this paper in the {\tt Enterprise}~\cite{enterprise} and  {\tt Discovery}~\cite{discovery} PTA data-analysis packages.
Most of our results were produced using {\tt Discovery}, exploiting its JAX-based \cite{jax2018github} autodifferentiation \cite{baydin2018automatic} to sample Bayesian posteriors using Hamiltonian Monte Carlo, in the ``No-U-Turn'' sampler implementation \cite{hoffman2014no} of \texttt{NumPyro} \cite{phan2019composable,bingham2019pyro}.

\section{Covariance matrices in PTA data analysis}
\label{sec:cov-matrix}

In PTA data analysis, the timing residuals $\tresid$ (i.e., the differences between the measured and expected TOAs, as predicted by a deterministic \emph{timing model}) are described as arising from a combination of white measurement noise and one or more time-correlated processes. In many instances the correlated processes (e.g., intrinsic pulsar spin noise or the common GWB signal) are assumed to be Gaussian and stationary. Although any complete basis can be used to develop their low-rank representation, traditionally these processes are described in either the time domain or the Fourier domain. In this section we summarize the theory of Gaussian-process covariance, we describe its standard Fourier-domain formulation, and we introduce our new 
method.

\subsection{Time-series modeling of stationary processes}
\label{ssec:timeseries}

Let $y(t)$ denote a real zero-mean stationary random process
\be
\langle y(t) \rangle = 0 \,,
\ee
with autocorrelation function~\cite{2009MNRAS.395.1005V}
\be
\langle y(t)\, y(t') \rangle \equiv C(|t-t'|) \,,
\label{2ptf_revised}
\ee
which depends solely on the time-lag $\tau=|t-t'|$. By the Wiener–Khinchin theorem, the autocorrelation function is the Fourier transform of the two-sided power spectral density (PSD) $\mathcal{S}(f)$:
\be
C(\tau) = \int_{-\infty}^{+\infty} \dd f\, \mathcal{S}(f) \, e^{i \, 2\pi f \tau} \,.
\label{WK_theorem}
\ee
The two-sided PSD $\mathcal{S}(f)$ is defined for $-\infty<f<+\infty$ with $\mathcal{S}(-f)=\mathcal{S}(f)$. In the Fourier domain the same process is represented as
\be
\langle \tilde{y}(f) \rangle = 0 \,,
\ee
with
\be
\langle \tilde{y}(f) \, \tilde{y}^{*}(f') \rangle = \mathcal{S}(f)\,\delta(f-f') \,,
\label{Phi-diag}
\ee
where the Fourier transform of $y(t)$ is defined in the usual way:
\be
\tilde{y}(f) = \int_{-\infty}^{+\infty} \dd t\, y(t) \, e^{-i2\pi f t} \,.
\ee

Expanding $y(t)$ in a complex Fourier series over the finite time interval $t\in[-T/2,T/2]$, we write
\be
y(t) = \sum_{k=-\infty}^{+\infty} \tilde{y}^k\, e^{i2\pi f_k t} \,,
\label{FourierExpansion}
\ee
where the discrete frequencies are given by $f_k=k/T$. The Fourier coefficients can be obtained via
\be
\tilde{y}^k = \frac{1}{T} \int_{-T/2}^{T/2} \dd t\, y(t)\, e^{-i2\pi f_k t}\,.
\label{FourierAmplitudes}
\ee

From these definitions the Fourier-domain covariance
(or the power in each Fourier bin)
follows as
\be
\begin{aligned}
S_{jk} \equiv& \langle \tilde{y}^j\, \tilde{y}^{k*}\rangle  \\
=&\frac{1}{T^2}\int_{-T/2}^{T/2}\!\dd t \int_{-T/2}^{T/2}\!\dd t'\, C(|t-t'|) \,e^{-i2\pi f_j t}\, e^{i2\pi f_k t'} \,.
\end{aligned}
\label{FourierCovariance}
\ee
Using the Wiener–Khinchin theorem with Eq.~(\ref{Phi-diag}), this double time integral can be re-expressed as an integral over frequency \cite{Allen:2024uqs}:
\be
S_{jk} = \int_{-\infty}^{+\infty} \!\dd f\, \mathcal{S}(f) \, \sinc\bigl[\pi T (f-f_j)\bigr]\, \sinc\bigl[\pi T (f-f_k)\bigr] \,,
\label{Phi-sinc}
\ee
where $\sinc(x)$ denotes the cardinal sine function $\sin(x)/x$. Equation~\eqref{Phi-sinc} demonstrates that, even for a stationary process, the Fourier coefficients defined over a finite interval are not strictly uncorrelated but rather display correlations mediated by the sinc functions. Now, in current PTA analyses the covariance is typically approximated in the diagonal form
\be
S_{jk} = \frac{\mathcal{S}(f_j)}{T}\, \delta_{jk}\,,
\label{Phi-diag_recap}
\ee
which neglects these inter-frequency correlations. In what follows we show that this diagonal approximation may affect spectral-parameter inference, motivating the development of improved computational methods that fully capture the frequency correlations.

\subsection{Standard method: Fourier sum}

PTA datasets consist of a large number of pulsar-timing residuals $\tresid$ observed at unevenly spaced times. The $\tresid$ are produced by subtracting a best-fit theoretical model from the TOAs \cite{2023ApJ...951L...9A}.
The residuals are modeled as the sum of a number of Gaussian noise processes: these may have independent probability distributions for each residual, or they may be correlated across TOAs (the former are usually called ``white noise,'' reserving the term ``Gaussian processes'' for the latter).
If a time-correlated Gaussian process, such as intrinsic spin noise or the GWB, is also stationary, it is fully described by the autocorrelation function of the residuals $C(\tau)$, which is related to the PSD by the Wiener--Khinchin theorem of Eq.~\eqref{WK_theorem}. In standard PTA analyses, it is customary to work with the one-sided PSD
\be
S(f) \;=\;
\begin{cases}
2\,\mathcal{S}(f), & f>0,\\
0, & f<0,
\end{cases}
\ee
so that all the power resides at positive frequencies. With this convention
\be
C(\tau) = \int_0^{+\infty} \dd f \, S(f) \cos(2\pi f \tau) \,. \label{cov}
\ee
In the rest of this paper we will work with the one-sided PSD $S(f)$.
If $S(f)$ diverges sufficiently strongly for $f \rightarrow 0$, as is the case for a power law $\propto f^{-\gamma}$ with $\gamma \geq 1$, the integral of Eq.\ \eqref{cov} is divergent, requiring the introduction of a low-frequency cutoff or regularization. For a generic power law, this results in a time-domain power law $\propto \tau^{\gamma - 1}$ corrected by a hypergeometric function \cite{2009MNRAS.395.1005V}. For power-law PSDs with $\gamma < 1$ the integral diverges as $f\rightarrow \infty$ and a high-frequency cutoff or regularization is needed.

The PTA log likelihood can be written in general as
\be
\begin{aligned}
    \log p(\tresid|\Lambda) = &-\frac{1}{2} (\tresid - M\bm{\epsilon})^T \bigl[N + C(\Lambda)\bigr]^{-1}(\tresid-M\bm{\epsilon}) \\ &- \frac{1}{2}\log \bigl|2 \pi\bigl[N + C(\Lambda)\bigr]\bigr|\,, \label{eq:original_covariance}
\end{aligned}
\ee%
where $M$ is a design matrix given by the derivatives of the deterministic model with respect to the model parameters (e.g., the spin period and period derivative of the neutron star), and the $\bm{\epsilon}$ are linear corrections to those parameters; $N$ is the covariance matrix of white (measurement) noise; $C$ is the sum of the Gaussian-process covariance matrices, which are functions of the noise hyperparameters $\Lambda$. As written here, $C$ is a large dense matrix, so computing its inverse is impractical for real datasets. Furthermore, the elements of $C$ may themselves be too expensive to compute or numerically unstable, as is the case for the hypergeometric function.

Instead, finite Gaussian-process bases are used to reduce the dimensionality of the problem and evaluate the likelihood more efficiently \cite{2009MNRAS.395.1005V,Lentati:2012xb,vanHaasteren:2012hj,2014PhRvD..90j4012V,vanHaasterenVallisneri2015,taylor2021nanohertz}.
To wit, Gaussian processes are written as the product of a basis $F$ and weights $\mathbf{a}$, which follow a Gaussian prior. So the PTA log likelihood may be written (omitting the timing model corrections for the moment)
\be
\begin{aligned}
    \log p(\tresid | \mathbf{a}) = &-\frac{1}{2}(\tresid - F\mathbf{a})^T N^{-1} (\tresid - F\mathbf{a}) \\
    &- \frac{1}{2}\log|2 \pi N| \,,\label{eq:reduced_rank_likelihood} 
\end{aligned}
\ee
with
\begin{equation}
\log p(\mathbf{a} | \Lambda) = -\frac{1}{2}\mathbf{a}^T \Phi^{-1}(\Lambda) \mathbf{a} - \frac{1}{2}\log| 2\pi \Phi(\Lambda)|\,,
\end{equation}
and marginalizing $p(\tresid | \mathbf{a}) p(\mathbf{a} | \Lambda)$ over the weights $\mathbf{a}$,
\be
\begin{aligned}
    \log p(\tresid | \Lambda) = &-\frac{1}{2} \tresid^T\bigl[N + F\Phi(\Lambda)F^T\bigr]^{-1}\tresid \\
    &- \frac{1}{2}\log \bigl|2 \pi\bigl[N + F\Phi(\Lambda)F^T\bigr]\bigr| \,.\label{eq:reduced_rank_posterior}
\end{aligned}
\ee
While the matrix $N + F \Phi F^T$ is still large, it can be inverted using the Woodbury lemma,
\be
\begin{aligned}
\bigl[N &+ F\Phi(\Lambda)F^T\bigr]^{-1} = N^{-1} \\[1ex]
&- N^{-1} F (\Phi^{-1} + F^T N^{-1} F)^{-1} F^T N^{-1} \,;
\label{woodbury}
\end{aligned}
\ee
this expression is much more computationally efficient since $N$ is usually diagonal (or almost), and the inner inverses in \eqref{woodbury} have the dimension of the number of bases, which is much smaller than the number of residuals (a few tens vs.\ tens of thousands). 
We can perform a similar marginalization process for the timing-model corrections $\bm{\epsilon}$, usually by adopting a broad, uninformative prior.
We discuss the effect of timing-model marginalization on the low-rank reconstruction of $C$ in Sec.~\ref{sec:comparison}.

Beginning with Ref.~\cite{Lentati:2012xb}, PTA analyses have used Fourier bases for this low-rank representation: the $\mathbf{a}$ are Fourier coefficients for frequencies $f_k = k/T$, with $T$ the total observation time and $k = 1, \ldots, n$; and $F$ is a discrete Fourier transform matrix of sines and cosines,
albeit evaluated at unequally spaced times. 
Comparing Eqs.\ (\ref{eq:original_covariance}) and (\ref{eq:reduced_rank_posterior}), we see that the full $C$ is approximated as $F\Phi F^T$.
Evaluating the Fourier basis at centered times, we have
\be
\begin{aligned}
C_{a b} \approx & \sum_{j,k=1}^n \, \Phi^{cc}_{j k} \cos(2\pi \, j\,t_a/T)\cos(2\pi\,k\,t_b/T) \\
& \quad\,\, + \, \Phi^{ss}_{j k} \sin(2\pi\,j\,t_a/T)\sin(2\pi\,k\,t_b/T) \,. \label{sinecosine}
\end{aligned}
\ee
where
\be
\Phi^{cc}_{jk}=2(S_{jk}+S_{j,-k})\,,
\quad
\Phi^{ss}_{jk}=2(S_{jk}-S_{j,-k})\,,
\label{eq:real_fourier_covariance}
\ee
and $S_{jk}$ is the covariance of Eq.~\eqref{Phi-sinc} evaluated for both positive and negative Fourier indices.
For the centered window used there, $S_{jk}$ is real and the sine--cosine covariance blocks vanish by reflection symmetry.

The main issue here is that, up through the recent GWB-evidence analyses \cite{NANOGrav:2023gor,EPTA:2023fyk,Reardon:2023gzh,MilesShannon2025,Xu:2023wog}, the Gaussian-process prior is set from Eq.~\eqref{Phi-diag_recap}: $\Phi^{cc}_{jk}=\Phi^{ss}_{jk} = S(j/T)/T \, \d_{j k}$, neglecting the inter-frequency and positive--negative-frequency correlations of Eq.\ \eqref{Phi-sinc}. In Sec.\ \ref{PTA} we will see that these correlations have measurable consequences on parameter estimation.
Furthermore, diagonal $\Phi_{j k}$ correspond to a periodic correlation function (with $C(0) = C(T)$), which is unphysical for the nonperiodic noise processes of interest, as noted already in~\cite{vanHaasterenVallisneri2015}.

While it may seem that this problem could be solved by using Eq.\ \eqref{sinecosine} and taking $\Phi_{j k}$ from Eq.\ \eqref{Phi-sinc}, doing so poses further problems.
First, Eq.\ \eqref{sinecosine} is then effectively a two-dimensional Fourier series over a finite domain, which will suffer from Gibbs ringing artifacts near its boundaries \cite{gibbs1898Natur..59..200G,wilbraham1848certain}, slowing the convergence of the expansion;
we will explore this numerically in Sec.\ \ref{sec:comparison}. 
Second, Eq.\ \eqref{Phi-sinc} requires performing a highly oscillatory integral that is numerically delicate.

\subsection{New method: Wiener--Khinchin by FFT + linear interpolation}
\label{ssec:new-method}

In this section we introduce a novel method to approximate the full covariance matrix efficiently and accurately without performing the computationally intensive double-sinc integral of Eq.~\eqref{Phi-sinc}. 
Our method relies on evaluating the time-domain autocorrelation function $C(\tau)$ on a coarse time grid using a fast Fourier transform of the PSD, thus retaining the frequency-correlation information that is lost in the diagonal approximation of Eq.~\eqref{Phi-diag_recap}.
The coarse-time covariance matrix is then interpolated efficiently to the full set of TOAs, with the additional advantage that the locally concentrated interpolation basis mitigates finite-window effects and artifacts arising from the periodic boundary conditions implicitly assumed in the Fourier-basis approach.

For a stationary Gaussian process, the time-domain covariance matrix for a set of regularly spaced times has a \emph{Toeplitz} structure where each descending diagonal is set to the value of the autocorrelation function for the corresponding lag.
That is, if $C_{a b} = C(|t_a - t_b|)$ with $t_a = a \Delta t$ and $a = 0, \ldots, N - 1$, then the main diagonal $C_{a a}$ would be set to $C(\tau = 0)$; the super- and sub-diagonals $C_{a,a \pm 1}$ would be set to $C(\tau = \Delta t)$, and so on.
This means that the full covariance matrix can be recovered from the vector $c_a = C(a \Delta t)$, which we will compute using the fast Fourier transform.

Furthermore, for noise processes dominated by the lowest frequencies (as are those of interest in PTA data analysis), the autocorrelation function varies smoothly, which makes it possible to represent it on a coarse time grid and then interpolate to every TOA.
The resulting covariance matrix can be written as $B \hat{C} B^T$, where $\hat{C}$ is the coarse-grid covariance matrix, and the interpolation matrix $B$ has a locally concentrated structure set by the interpolation order.
For linear interpolation, which we use throughout unless otherwise specified, the entries of $B$ are given by the piecewise-linear (``hat'') basis functions associated with the coarse grid, evaluated at the TOAs. Equivalently, if a TOA $t_a$ lies between two neighboring coarse-grid points $\hat t_b$ and $\hat t_{b+1}$, it is represented as a linear combination of those two points with weights determined by the relative distances to them; see, e.g., \cite{strang1986}.
This $B \hat{C} B^T$ can be inserted directly in Eq.\ \eqref{eq:reduced_rank_posterior} as a replacement for $F \Phi F^T$.

Finally, we compute the coarse-time autocorrelation vector $\hat{c}_a$ by evaluating the Wiener--Khinchin integral \eqref{cov} as the Fourier sum
\be
\begin{aligned}
\hat{c}_a & \approx \Delta f \sum_{k=0}^{n_\mathrm{max}-1} S(f_k) \cos(2 \pi f_k \tau_a)\\
& = \Delta f \sum_{k=0}^{n_\mathrm{max}-1} S(k \, \Delta f) \cos(2\pi \, a \, k \, \Delta \hat{t} \, \Delta f)\,,
\label{iffteq}
\end{aligned}
\ee
where $\Delta f = 1 / (\omega T)$, with $\omega$ an integer \emph{oversampling factor}; $\Delta \hat{t}$ is the spacing of coarse times; and $n$ is chosen such that $n_\mathrm{max} \Delta f \approx \z / (2 \Delta \hat{t})$ (i.e., we integrate to $\z$ times the Nyquist frequency of the coarse time grid).
Setting $\omega > 1$ increases the frequency resolution, while $\z > 1$ ensures that higher frequencies are included in the Wiener--Khinchin integral.

We evaluate Eq.\ \eqref{iffteq} using the inverse fast cosine transform, which results in a vector of length $2 \z \omega n$; $\hat{c}_a$ is then given by every $\z$-th element of the transform, for the first $\hat N = 2n + 1$ elements.\footnote{For efficiency, our implementation requires $\omega > 1$.}
This procedure ensures the accurate and efficient evaluation of the Wiener--Khinchin integral and allows for a controlled trade-off between computational cost and accuracy by choosing $\omega$ and $\z$.
When analyzing an array of pulsars together, the same coarse time basis and covariance matrix can be shared among them, as long as $T$ is set to the total time span of the multipulsar dataset, and the appropriate matrix $B_i$ is used for each pulsar.

Chromatic processes, such as dispersion-measure variations, use the same construction. The stationary process on the coarse grid is the latent dispersion measure: $\hat{C}$ is its Toeplitz covariance, obtained from the Wiener--Khinchin transform of the dispersion-measure spectrum. The delay in a timing residual is proportional to $(\nu/\nu_{\mathrm{ref}})^{-2}$, and that factor multiplies the corresponding row of $B$, just as it multiplies a Fourier basis vector in the standard Gaussian-process model. Each TOA remains a linear combination of the two neighboring coarse-grid nodes; only the weight of that row includes the observing frequency. A general chromatic process is identical, with $(\nu/\nu_{\mathrm{ref}})^{-2}$ replaced by $(\nu/\nu_{\mathrm{ref}})^{-\alpha}$, where $\alpha$ is the chromatic index and $\nu_{\mathrm{ref}}$ a reference frequency.

The same interpolation applies to irregular sampling and to gaps, whether or not the process is chromatic. The coarse grid is regular across the full span, gaps included, and $\hat{C}$ is built on that grid alone. Each TOA is interpolated only between the two coarse-grid nodes that bracket it, one spacing $\Delta\hat{t}$ apart. Nodes that fall inside a gap have no nearby TOA, but they remain part of the latent process, and the correlation across the gap is carried by $\hat{C}$ at that lag.

\section{Comparison of methods}
\label{sec:comparison}

In this section, we compare the true full-rank time-domain covariance matrix with the three low-rank reconstructions discussed above.

\subsection{Comparing covariance matrices for Mat\'ern covariance function}
\label{sec:comparison_matern}

For this comparison, we study a numerically well-behaved PSD that is low-frequency--dominated and sufficiently similar to the signals that appear in PTA models.
Namely, we focus on a \emph{Mat\'ern} process of order 3/2, which has PSD
\be
S_{3/2} (f) = \frac{24 \, \sqrt{3} \, \lambda  \, \sigma ^2}{\left[\left(2 \pi \lambda f \right)^2 + 3 \right]^2} \,, \label{SMatern}
\ee
where $\lambda$ is the length scale of the process and $\sigma$ the variance.
Note that $S_{3/2}(f)$ goes to a constant as $f \rightarrow 0$ and vanishes as $f^{-4}$ for $f \rightarrow \infty$. Thus $C(\tau)$ is always well defined.
The corresponding correlation function is given by integrating $S_{3/2}(f)$ in Eq.\ \eqref{cov}:
\be
C_{3/2}(\tau) = \frac{\sigma^2 \left(\lambda +\sqrt{3} \tau \right) e^{-\frac{\sqrt{3} \tau }{\lambda }}}{\lambda } \,. \label{CMatern}
\ee
For easy reference, we name the four methods under consideration:
\begin{itemize}
\item \TC: the exact covariance matrix given by Eq.\ \eqref{CMatern};
\item \DP: the low-rank approximation $F \Phi F^T$ with diagonal $\Phi$ set from Eqs.\ \eqref{Phi-diag_recap} and \eqref{SMatern};
\item \SP: the low-rank approximation $F \Phi F^T$ with a non-diagonal $\Phi$ computed from Eqs.\ \eqref{Phi-sinc} and \eqref{SMatern};
\item \FFT: our new approximation $B \hat C B^T$.
\end{itemize}

\begin{figure*}
    \begin{center}
    \includegraphics[width=6in]{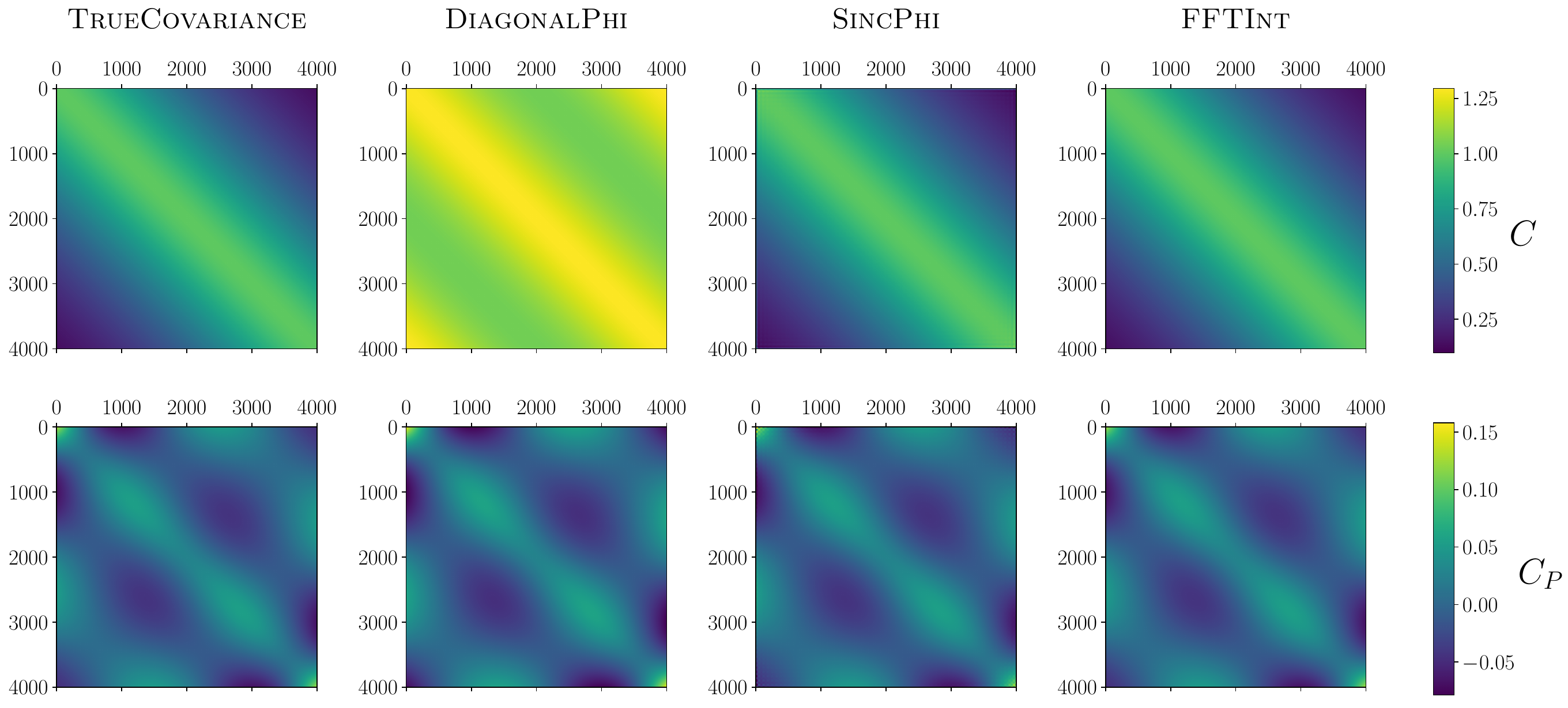}
    \end{center} 
    \caption{\textbf{Top}: $\text{Mat\'ern}_{3/2}$ covariance matrices according to the \TC\ (exact), \DP, \SP, and \FFT\ methods. The descending-diagonal Toeplitz structure is evident for all. \DP\ stands apart because the diagonal kernel makes it periodic along the minor diagonal.
    \textbf{Bottom}: quadratic-projected covariance matrices $C_P$. The projection creates an obvious multipolar structure and reduces the differences between the four matrices. Note the reduced range of the color scale.}
    \label{fig:cov-matrices}
\end{figure*}

In the top row of Fig.\ \ref{fig:cov-matrices} we show these covariance matrices for an evenly spaced series of $2\,001$ observation times over the interval $[2\,000, 6\,000]$ with duration $T = 4\,000$ (in arbitrary dimensionless units).
We set $\l=2\,000$ and $\s=1$; we take $n = 60$ for \DP\ and \SP; and, to cover the same frequency band, we use $\hat N=121$ coarse-grid nodes for \FFT, with $\o=6$ and $\z=1$, for a total of $361$ FFT frequencies.
This comparison shows that \DP\ is a very inaccurate approximation to \TC, even for the main diagonal (which is equal to $C(0)$), while \SP\ and \FFT\ appear to be much closer. This happens because for \DP\ $C(0) = T^{-1} \sum_k S(f_k)$, while \SP\ and \FFT\ embody more accurate representations of the Wiener--Khinchin integral that can access the value of $S(f)$ between ``bin'' frequencies $k/T$.

However, comparing the matrices in this fashion is not entirely relevant to PTA data analysis.
In the bottom row of Fig.\ \ref{fig:cov-matrices} we show the same covariance matrices after projecting out the subspace of residuals orthogonal to the family of generic quadratic functions of the TOAs.
Doing so mimics the most significant effects of marginalizing the PTA likelihood over timing-model errors, which is standard practice for PTAs \cite{vanHaasteren:2012hj}.
The projected covariance matrix is $C_{P} \equiv P C P^T$, where the projector $P = I - M (M^T M)^{-1} M^T$, and $M$ is a \emph{design matrix} with columns given by constant, linear, and quadratic functions of the TOAs.
The projection reduces differences significantly. This can be partly understood by observing that the Fourier basis is strictly periodic with period $T$, which means that the first and last observations in the time-domain have the same response. The presence of a constant and linear basis in $M$ breaks that symmetry, and allows the approximation to more accurately follow the model.

\begin{figure*}
    \begin{center}
    \includegraphics[width=6in]{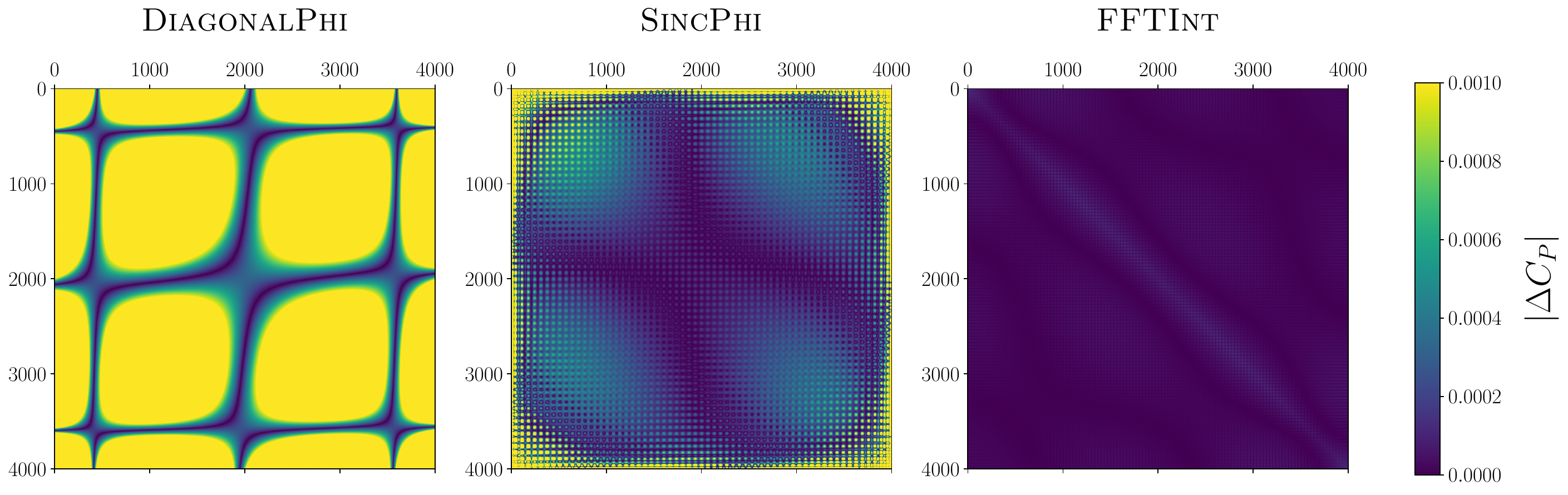}
    \end{center}
    \caption{Element-wise absolute difference between the exact and low-rank $\text{Mat\'ern}_{3/2}$ covariance matrices, shown here after projecting out generic quadratic functions. Note that the ``pixelated'' structure of the \SP\ method is real: it is a manifestation of the Gibbs phenomenon that can also be seen in Fig.~\ref{fig:signal-rec}, and it is more pronounced at the edges of the time window.}
    \label{fig:cov-matrices-diff}
\end{figure*}

The absolute element-wise differences $|\Delta C|$ between \TC\ and the approximated covariances are plotted in Fig.~\ref{fig:cov-matrices-diff} (post projection), and summarized in Table \ref{tab:l1_comparison}, which shows the matrix-averaged difference pre and post projection.  Although the four matrices at the bottom of Fig.~\ref{fig:cov-matrices} look very similar, the \DP\ errors remain much larger, and \SP\ gets in trouble at the edges of the matrix, with the Gibbs phenomenon apparent in the dense regular pattern of high-error ``dots.'' By contrast, \FFT\ remains very accurate across all TOA pairs.

In Sec.~\ref{sec:stresstest} we repeat the comparison for a family of Gaussian PSDs. These form an over-complete basis, so a broad class of smooth spectra can be written as sums of them. The low-frequency--dominated members behave as the Mat\'ern process does here, whereas narrow in-band features expose the failure of the diagonal approximation, which then has errors of order unity even after the timing-model projection.
\begin{table}
    \centering
    \setlength{\tabcolsep}{15pt}
    \renewcommand{\arraystretch}{1.2}
    \begin{tabular}{c|c|c}
        Reconstruction & $\overline{|\Delta C|}$ & $\overline{|\Delta C_P|}$ \\
        \hline
        \DP & $0.46$ & $2.3\cdot 10^{-3}$ \\
        \SP & $4\cdot 10^{-3}$ & $5.5\cdot 10^{-4}$ \\
        \FFT & $3\cdot 10^{-5}$ & $1.8\cdot 10^{-5}$ \\
    \end{tabular}
    \caption{Average element-wise absolute difference between the exact and low-rank $\text{Mat\'ern}_{3/2}$ covariance matrices, shown pre ($\overline{|\Delta C|}$) and post ($\overline{|\Delta C_P|}$) quadratic projection.}
    \label{tab:l1_comparison}
\end{table}

\subsection{Convergence of the covariance function}
\label{sec:cov_convergence}

Next, we are interested in how the low-rank approximation of the covariance matrix converges to the true covariance matrix as a function of approximation parameters $\hat{N}$, $\omega$, and $\z$.
As in Sec.~\ref{sec:comparison_matern}, we focus on the Mat\'ern PSD, but we also study a pure power-law PSD.

In the PTA literature, power-law signals are typically parameterized as
\be
\label{SPL}
S_{\text{PL}}(f) = \frac{A^2}{12\pi^2 f_{\text{ref}}^3} \left( \frac{f}{f_{\text{ref}}} \right)^{-\g} \,,
\ee
where $f_{\text{ref}} = 1 \ \text{yr}^{-1}$, $A$ is the amplitude of the noise process at $f_{\text{ref}}$, and $\gamma$ is the positive spectral index.
Power-law signals are problematic from a theoretical point of view, because the integrated variance of the unregularized power law is infinite: for $\g \geq 1$ the power diverges as $f\rightarrow 0$, while for $\g \leq 1$ the total power diverges as $f\rightarrow \infty$. Nevertheless, power-law signals are still used in PTA practice, because the timing model absorbs the $f=0$ divergence when $1 < \g < 7$, while the low-rank approximation sets a high-frequency cutoff. It remains unphysical to consider $\gamma \in [0,1]$ or $[7,10]$.

For our convergence study, it is necessary to incorporate a ``break'' in the power law (i.e., a change in spectral slope) so that we avoid the divergence of total power. This can be done as follows:
\be
\label{SBrokenPL}
S_{\text{BPL}}(f) = S_{\text{PL}}(f) \left[ 1+\left( \frac{f}{f_b}\right)^\frac{1}{\k}\right]^{\k (\g - \d)} \,,
\ee
where $\k$ parameterizes the smoothness of the transition, $f_b$ is the break frequency, and $\gamma$ (or $\d$) is the spectral slope when $f$ is sufficiently below (above) $f_b$.
For this convergence study we set $\kappa = 1/2$, $f_b / f_\mathrm{ref} = 1 / 50$, $\g=0$, and $\d=13/3$. The choice of $\g=0$ causes the PSD to be flat for $f\rightarrow 0$, while $\d=13/3$ sets the slope to a typical steep power law for $f > f_b$. For the Mat\'ern PSD of Eq.~\eqref{SMatern} we set $\lambda=T$. The amplitudes $\sigma$ and $A$ for the Mat\'ern process and the broken power law are set so that \TC\ has unit diagonal.

\begin{figure*}
    \begin{center}
    \includegraphics[width=6in]{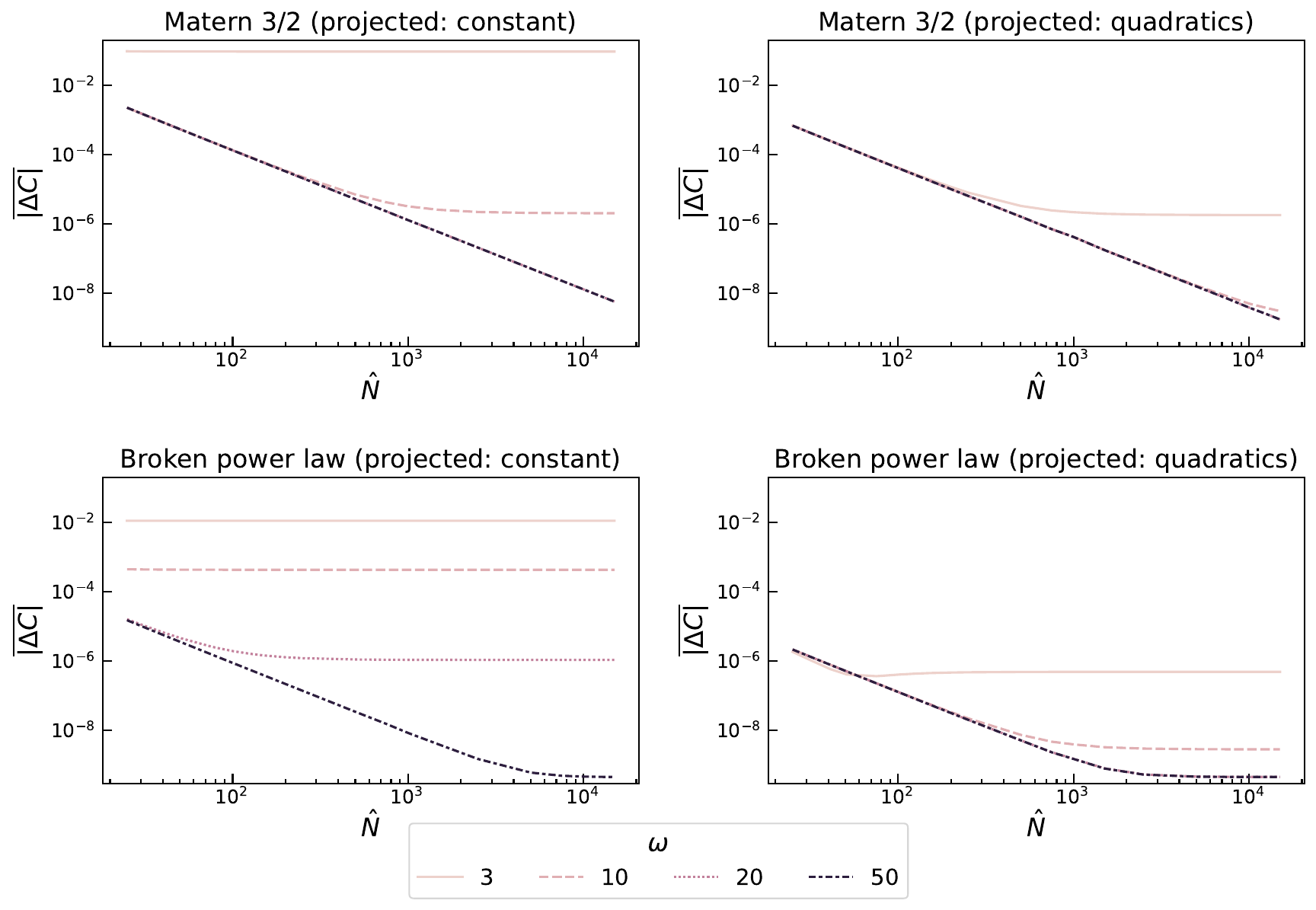}
    \end{center} 
    \caption{Mean absolute element-wise difference of the approximated and true covariance matrix as a function of the number of coarse timestamps $\hat N$ and the oversampling factor $\omega$ for the Matérn$_{3/2}$ and broken power-law kernels. All amplitudes are normalized such that the true (unprojected) covariance matrix is unity on the diagonal. \\
    \textbf{Top left}: Matérn$_{3/2}$ kernel [Eq.~(\ref{CMatern}) with $\lambda=T$]. We project out a constant TOA reference time. \\
    \textbf{Top right}: Same as top left, except we project out a generic quadratic function. \\
    \textbf{Bottom left}: Same as top left, but for the broken power law of Eq.~\eqref{SBrokenPL}, with $\kappa = 1/2$, $f_b=f_\text{ref}/50$, $\g=0$, $\d=13/3$. We project out a constant TOA reference time. \\
    \textbf{Bottom right}: Same as bottom left, except we project out a generic quadratic function.
    }
    \label{fig:convergence}
\end{figure*}

In Fig.\ \ref{fig:convergence} we explore how \FFT\ covariance matrices converge as we increase $\hat{N}$ and $\omega$. For the comparisons on the left we project out a constant\footnote{This is reasonable since the absolute time reference of TOAs is always arbitrary in PTA data analysis.}, while on the right we project out a generic quadratic.
We find that $\z$ has a very small effect (except when $\hat{N}$ is obviously too small), so we always set $\z = 1$.
Convergence is approximately quadratic in $\hat{N}$ until it saturates at a value that decreases with $\omega$; convergence is also helped by the quadratic projection.

\subsection{Rank-Reduced Signal Reconstruction and Upsampling}\label{ssec:signal-rec}

\begin{figure*}
    \begin{center}
    \includegraphics[width=6in]{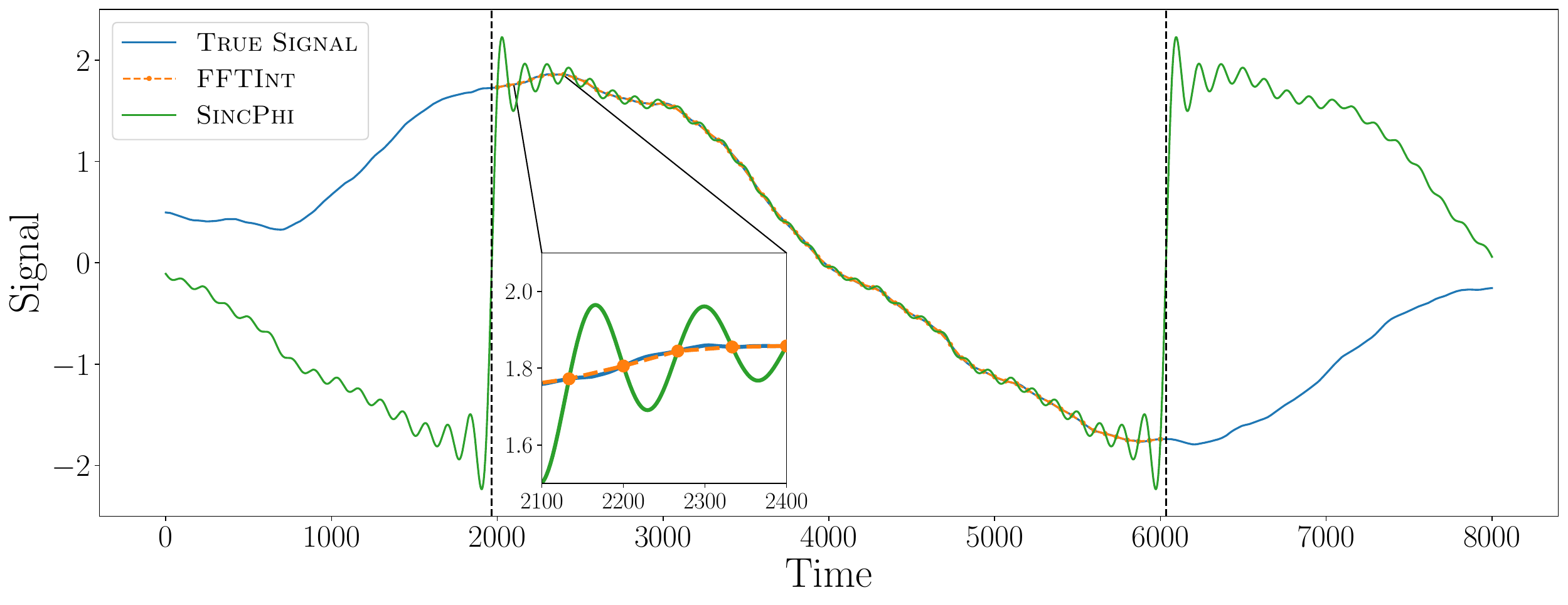}
    \end{center}
    \caption{\small Comparison of signal reconstruction with time- and frequency-domain bases. The blue curve depicts the ``true'' stochastic signal generated on a fine grid with $10\,000$ points over the interval $x \in [0,8\,000]$, with an observation window $x \in [2\,000,6\,000]$.
    The signal, as subsampled on a reduced set of $\hat N = 61$ evenly spaced node points (orange dots), with first observation at $x=2\,000$ and last observation at $x=6\,000$, is represented exactly in both the \SP\ Fourier and \FFT\ interpolation bases. When upsampled to the fine grid, the reconstruction via the interpolation basis (orange dashed line) recovers the true signal accurately, whereas the Fourier-based reconstruction (green curve) exhibits significant ringing due to the Gibbs phenomenon. The inset highlights this ringing effect near the boundaries of the observation interval. The vertical dashed lines, half a timestep before $x=2\,000$ and after $x=6\,000$, delimit the periodicity implied by the Fourier transform. See Appendix~\ref{app:ortho-fourier} for an explanation why the periodicity is not exactly equal to the observation window. 
    The \FFT\ method provides a local time-domain upsampling procedure that avoids the spurious oscillations associated with the Fourier approach.}
    \label{fig:signal-rec}
\end{figure*}

Figure~\ref{fig:signal-rec} illustrates the impact of the two rank-reduced representations on the reconstruction of a stochastic low-frequency signal, and highlights the differences that arise when we upsample the reduced representation. In this example, we begin with a \emph{true} signal generated on a finely sampled grid comprising $N=10\,000$ evenly spaced time points over the interval $x \in [0, 8\,000]$.
This smoothly varying waveform is plotted in blue. We then define the \emph{observation window} as the interval $x_{\rm obs} \in [2\,000,\, 6\,000]$,
which is indicated by vertical dashed lines in the figure. Although the underlying physical process extends over the full fine grid, we restrict PTA observations to this sub-interval; this distinction becomes crucial when comparing different reconstruction methods.

For the rank reduction we adopt both the FFT-Interpolation (\FFT) and the Fourier (\SP) methods. In our implementation we select $\hat N=61$ regularly spaced nodes within the observation window. This choice corresponds to a reduced Fourier basis with $n=30$ frequency modes. On the coarse-grained grid, the basis is complete---meaning that any signal defined on the nodes can be represented exactly either in the time domain or in the Fourier domain (via an invertible Fourier transform). The ``data'' in this reduced basis is the true signal subsampled on the node positions $\hat x$ to obtain the corresponding values $\hat y$; these values (shown as orange dots in the figure) are equivalently represented in the Fourier domain by the coefficients $a_{\rm SincPhi}$. Importantly, on the node grid the \FFT\ and the non-diagonal Fourier \SP\ approaches yield identical results: they cross at the orange dots. We use this one-dimensional signal reconstruction rather than a two-dimensional covariance comparison to illustrate differences between the methods more clearly. The same principle applies to the covariance matrices; see Eq.~(\ref{FourierCovariance}).

Once the reduced representation is fixed, we upsample the signal back to the full fine grid. In our \FFT\ approach the reconstruction is carried out in the time domain by applying the locally concentrated interpolation matrix $B$ to the coarse signal, yielding an orange dashed line that follows the original blue signal with high fidelity and without artifacts. In contrast, the upsampling in the Fourier representation proceeds by defining the Fourier basis over the same frequencies derived from $\hat x$ but evaluating the basis functions on the full fine grid. This procedure leads to the green curve in Fig.\ \ref{fig:signal-rec}. Owing to the inherent non-locality of the Fourier basis, the green reconstruction exhibits the well-known Gibbs phenomenon---a ringing artifact that is particularly pronounced at the edges of the observation window and continues over the entire domain. Interestingly, the Gibbs phenomenon and apparent edge discontinuity shows up also when we draw realizations from the \SP\ model.

The inset in Fig.\ \ref{fig:signal-rec} zooms into a representative region of the signal, clearly showcasing the ringing effect in the Fourier-based reconstruction. This example illustrates that, while both representations are exact on the reduced node grid, the \FFT\ method provides a local time-domain upsampling procedure that avoids the spurious oscillations associated with the Fourier approach.

The poor performance of \SP\ at the extremes of the dataset is clearly visible also in Fig.\ \ref{fig:cov-matrices-diff} where the absolute difference is maximum at the edges of the covariance matrix. 
Note that while the Gibbs phenomenon does not affect the coarse time basis, it will definitely affect the unevenly spaced TOAs found in real PTA datasets.

\subsection{Gaussian PSD stress test}
\label{sec:stresstest}
The family of Gaussians parameterized by amplitude, mean, and width forms an over-complete basis for the decomposition of smooth functions. This property of Gaussians is frequently used in many numerical applications, such as the use of Gaussian kernels in kernel density estimation. The attractive property of Gaussians for our purposes is that a Gaussian PSD has an analytical time-domain covariance function, which we exploit to set up a stress test and inspect how our method performs in this subsection.

\begin{figure*}
    \begin{center}
    \includegraphics[width=7in]{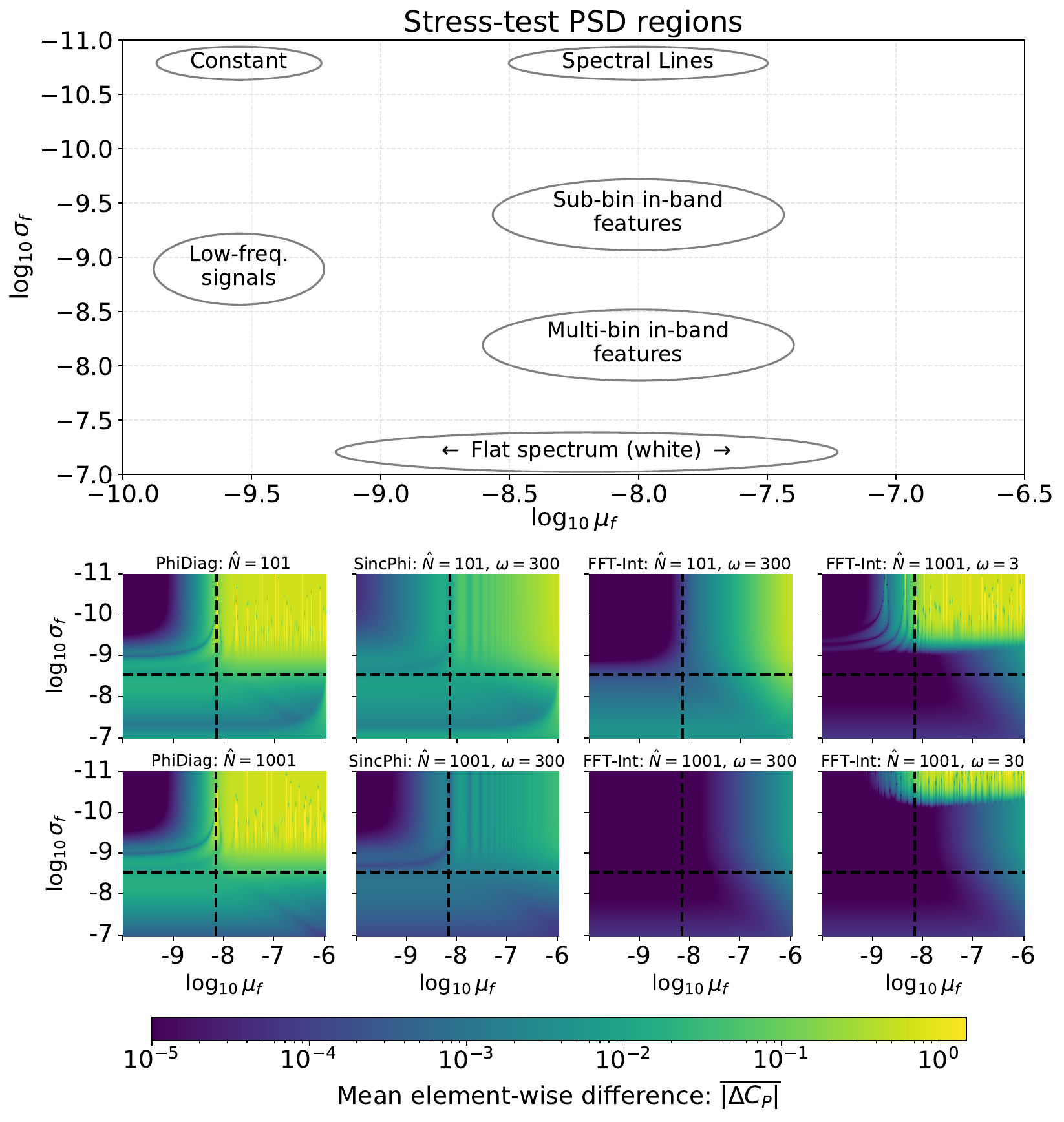}
    \end{center}
    \caption{Mean absolute element-wise projected difference $\overline{|\Delta C_P|}$ for various approximations of the covariance matrix.
    {\bf Top panel:} power spectral densities represented by a Gaussian of mean $\mu_f$ and variance $\sigma_f^2$.
    {\bf Bottom panels:} approximation error $\overline{|\Delta C_P|}$ for
    \DP, \SP, and \FFT\ 
    as a function of $\mu_f$ and $\sigma_f$.
    Horizontal and vertical dashed lines mark $f=1/T$.}
    \label{fig:stresstest}
\end{figure*}

We define the Gaussian one-sided PSD as:
\begin{equation}
    S(f) = \frac{A^2}{\sqrt{2\pi}\sigma_f} \left[ e^{-\frac{(f-\mu_f)^2}{2\sigma_f^2}} + e^{-\frac{(f+\mu_f)^2}{2\sigma_f^2}} \right] \,,
\end{equation}
where $f \geq 0$. We set amplitude $A=1$ throughout this subsection, and the mean and standard deviation of the Gaussian PSD peak are respectively $\mu_f$ and $\sigma_f$. The second term folds the mirror-image peak at $-\mu_f$ onto positive frequencies, so that $S(f)$ is the one-sided counterpart of a symmetric two-sided PSD. The corresponding covariance function, obtained from Eq.~\eqref{cov}, is:
\begin{equation}
    C(\tau) = A^2 \, e^{-2\pi^2\sigma_f^2\tau^2} \, \cos\left(2\pi\mu_f\tau\right).
\end{equation}
In Fig.~\ref{fig:stresstest} we show how the various approximations presented in this paper perform on this stress test. The top panel labels representative PSD regimes in the $(\mu_f, \sigma_f)$ plane: the low-frequency dominated PSDs are on the left, and the spectral lines are at the top. In the middle of the plot are features at some frequency spanning one or several bins.
The bottom panels show the average error of each method.
The \DP\ method breaks down for spectral lines and other in-band features narrower than the frequency-bin width $1/T$, with errors of order unity for both values of $\hat N$: a diagonal Fourier-domain prior cannot represent power that falls between the bin frequencies $k/T$.
The \SP\ method captures these features considerably better, but its accuracy remains limited, especially for $\hat N = 101$.
The \FFT\ method is the most accurate approximation across the plane, provided that the frequency resolution $1/(\omega T)$ of the Wiener--Khinchin sum resolves the width of the feature: for $\omega = 3$ and $30$ it fails for the narrowest lines, whereas for $\omega = 300$ and $\hat N = 1001$ its typical error is below $10^{-3}$.
Increasing $\hat N$ improves all methods for broad (white) spectra and for features approaching the Nyquist frequency of the coarse grid.
For low-frequency--dominated PSDs, which are the most relevant for PTA red noise, all methods remain reasonably accurate after projection, and \FFT\ is more accurate than the other methods by one to several orders of magnitude.
We conclude that the \DP\ approximation is adequate for smooth, low-frequency--dominated spectra, but not for spectra with sharp features.

\section{PTA analysis}
\label{PTA}

In this section we compare our new method (\FFT) to the standard diagonal-covariance formulation (\DP) in the analysis of a simulated dataset inspired by the NANOGrav 15-year data \cite{2023ApJ...951L...9A}.
The simulation is obtained by setting the pulsar white-noise parameters as in the NANOGrav dataset, and the red-noise parameters in the vicinity of their maximum-likelihood values in the NANOGrav data release. Red noise is generated in the time domain using the \FFT\ covariance matrix to color and correlate independent Gaussian samples:
we adopt a coarse grid of $\hat N = 241$ nodes, which yields a Nyquist frequency of $120/T$ Hz, where $T$ is the time span of the dataset ($\sim 16$ years). The oversampling factor $\omega$ is 100. This high-resolution construction is expected to approximate the target \TC\ accurately, although we do not quantify the residual error. Because \FFT\ is also used as a recovery model, the comparison should be understood as a convergence and consistency test rather than as a fully independent validation.
We model each pulsar's intrinsic spin noise with the two-parameter power law of Eq.~\eqref{SPL}. We also include a common spatially uncorrelated red-noise process (which represents the GWB), which follows the broken power law of Eq.~\eqref{SBrokenPL}, with $\k=0.1$.
For both functions, we enforce a low-frequency cutoff at $1/(10 \, T)$ Hz.

In the NANOGrav 15-year GWB analysis \cite{NANOGrav:2023gor,2024PhRvD.109j3012J}, this model was used with $\delta = 0$ (a flat tail) to determine the break frequency $f_b$ where red noise is overtaken by white measurement noise. Continuing the red-noise power law beyond that break leads to strong degeneracy between white- and red-noise parameters.
In the main NANOGrav analysis, spatially correlated red noise was then modeled with a simple power law with a hard cut at $f_b$. Both prescriptions lead to the same GWB posteriors. In this article, however, we wish to explore the impact of covariance-matrix approximations on the determination of $f_b$, so we adopt Eq.~\eqref{SBrokenPL} with $\delta = 1$---a mildly sloping tail to reduce the injection of power at higher frequencies.

\begin{figure*}
    \begin{center}
    \includegraphics[width=2.3in]{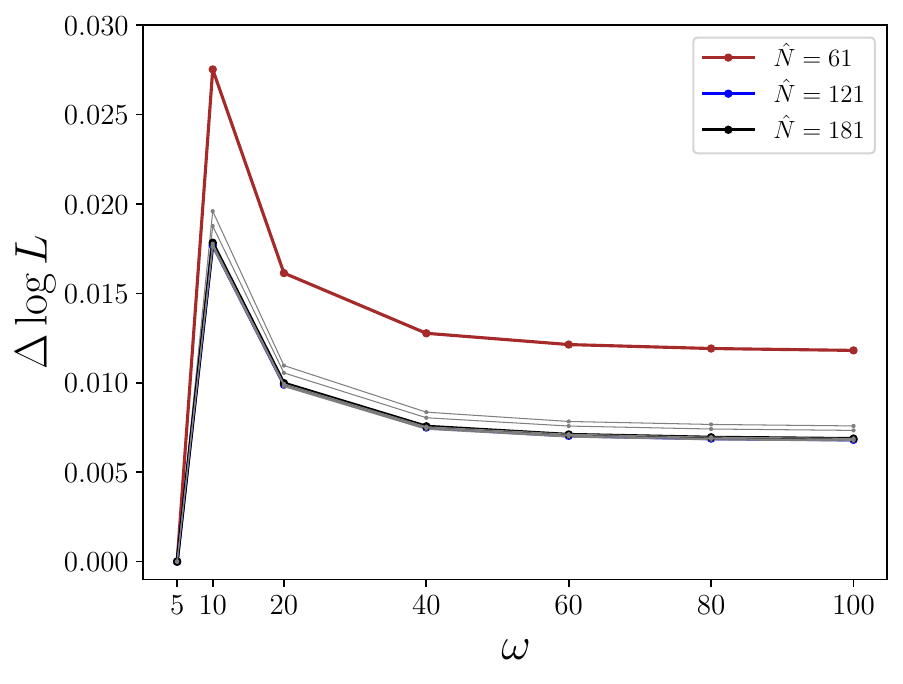}
    \includegraphics[width=2.3in]{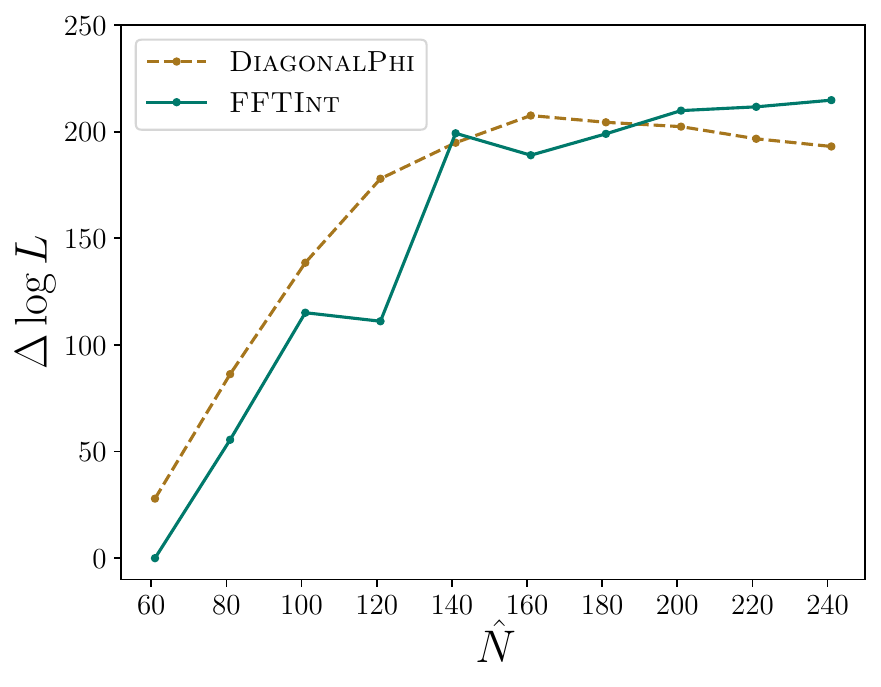}
    \includegraphics[width=2.3in]{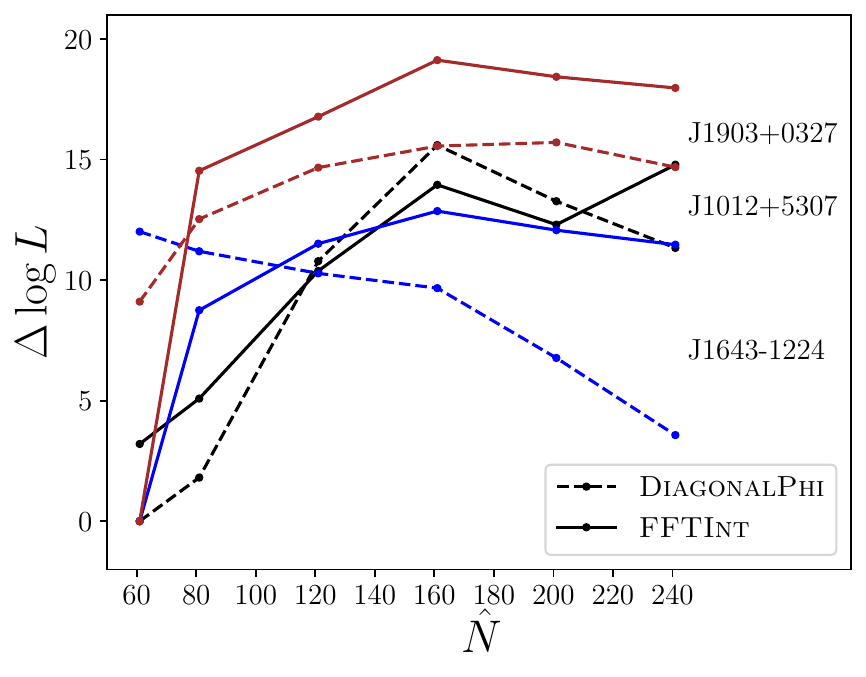}
    \end{center}
    \caption{Change $\Delta \log L$ (see the definition in the text) in the PTA likelihood computed using \FFT\ and \DP. {\bf Left}: change with $\omega$ and $\hat{N}$, for \FFT\ and the full NANOGrav array. {\bf Middle}: change with $\hat{N}$ for \FFT\ (with $\omega = 10$) and \DP\, for the full NANOGrav array. {\bf Right}: change with $\hat{N}$ for the $\Delta \log L$ contributions of three shallow-spectrum pulsars.}
    \label{fig:convergence-pta}
\end{figure*}

We sample $\log_{10} A_p$ and $\g_p$ for each pulsar $p$ and $\log_{10} A_\mathrm{gw}$, $\g_\mathrm{gw}$ and $\log_{10} f_b$ for the common process.
For a dataset of 67 pulsars this gives a total of 137 parameters to estimate.
We first explore how the likelihood, evaluated at the injected parameters, converges as we increase $\hat{N}$ and $\omega$ for the \FFT\ method and the number of frequency bins $n$ for the \DP\ method. For the same frequency content, $\hat{N}$ and $n$ are related by $\hat{N} = 2n+1$, so when we study the \DP\ likelihood as a function of $\hat N$, we are computing it for $n=(\hat N - 1)/2$.
In Fig.~\ref{fig:convergence-pta} we show the difference $\Delta \log L$
for the full PTA likelihood (left and middle) and for three pulsars (right) as we vary $\omega$ (for \FFT) and $\hat{N}$. Differences are computed with respect to \FFT\ with $\omega = 5$ (left), and to the minimum between \FFT\ and \DP\ at $\hat{N} = 61$ (middle and right).
These reference values are chosen only for plotting convenience and have no statistical significance.
Since $\Delta \log L$ is essentially constant for $\omega > 5$, we use $\omega = 10$ in the middle and right plots (and all subsequent analyses).
In the central panel we show how each likelihood converges with increasing $\hat N$, while retaining a method-dependent offset: both reach a plateau at $\hat N \simeq 140$. The variations for the larger $\hat{N}$ seem to be explained by shallow-spectrum pulsars ($ 1< \gamma < 2$), such as those shown in the right panel. In contrast, red-spectrum pulsars ($\gamma > 2$) are modeled equally well by both covariance matrices and have similar likelihoods.
Nevertheless, it appears that relatively large $\hat{N}$ may be needed to fully capture the variation of the NANOGrav PTA.

\begin{figure*}
    \begin{center}
    \includegraphics[width=7in]{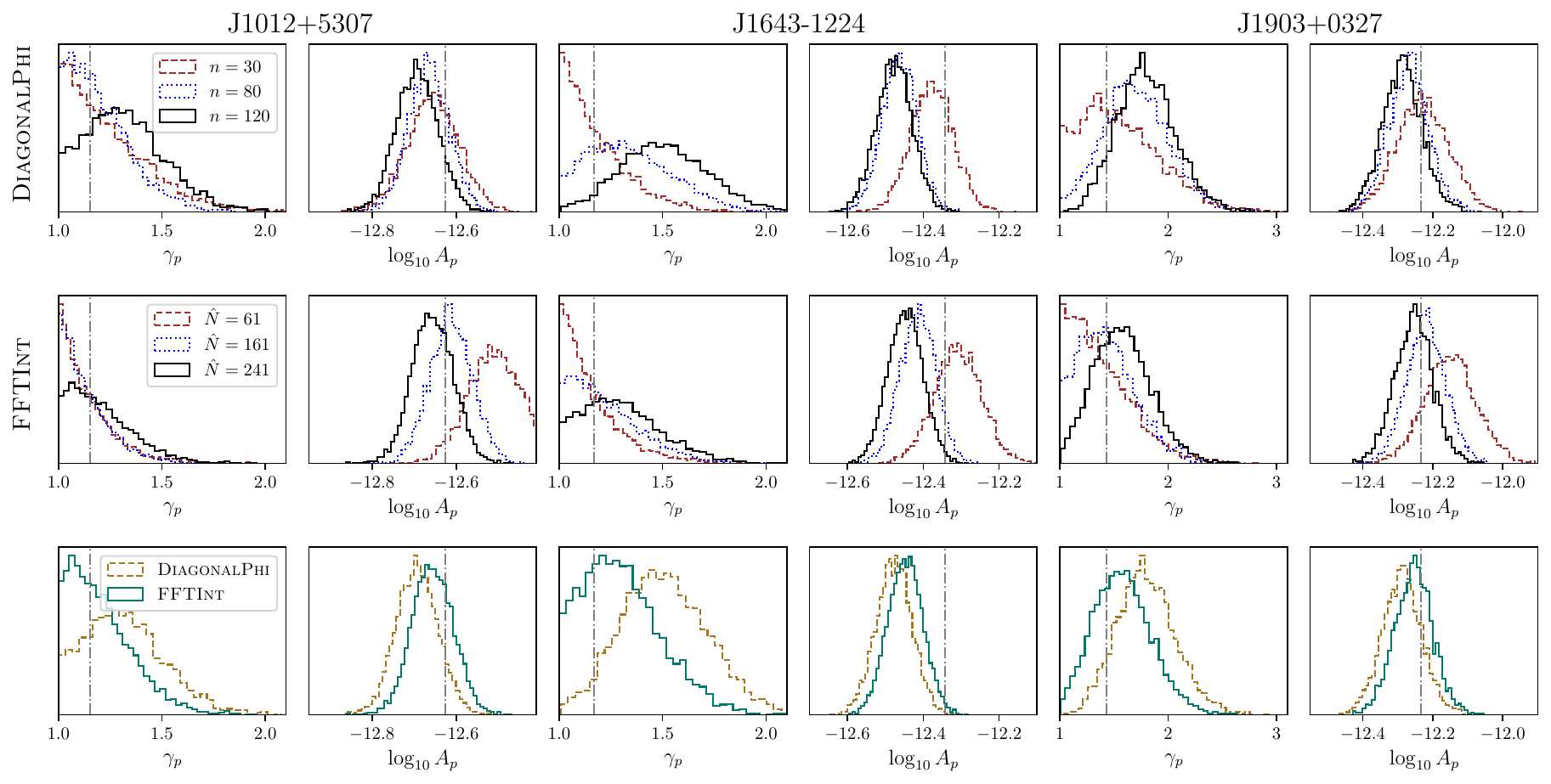}
    \end{center}
    \caption{Intrinsic timing-noise posteriors ($\gamma_p$ and $\log_{10} A_p$) for three simulated NANOGrav pulsars with shallow timing-noise spectra. The first row shows \DP\ posteriors with $n = 30, 80, 120$, the second row \FFT\ posteriors with $\hat{N} = 61, 161, 241$, and the third row compares the two methods for $\hat N = 241$ ($n=120$). Dashed--dotted lines show injected values.}
    \label{fig:pulsar}
\end{figure*}

In Fig.~\ref{fig:pulsar} we show the posteriors for the intrinsic red-noise parameters of the same three shallow-spectrum pulsars, as estimated as part of the full PTA analysis.
The first two rows explore the convergence of posteriors with increasing $\hat{N}$ for \DP\ and \FFT. The third row compares the two approximations with $\hat{N} = 241$ ($n = 120$).
Counterintuitively, \DP\ recovers the injected parameters more accurately with the smallest number of frequency bins than with larger values. As expected, \FFT\ progressively improves the inference accuracy as the number of coarse-time nodes increases.
For pulsars with red spectra, the two methods instead yield nearly identical posteriors for the intrinsic red-noise parameters.

\begin{figure*}
    \begin{center}
    \includegraphics[width=7in]{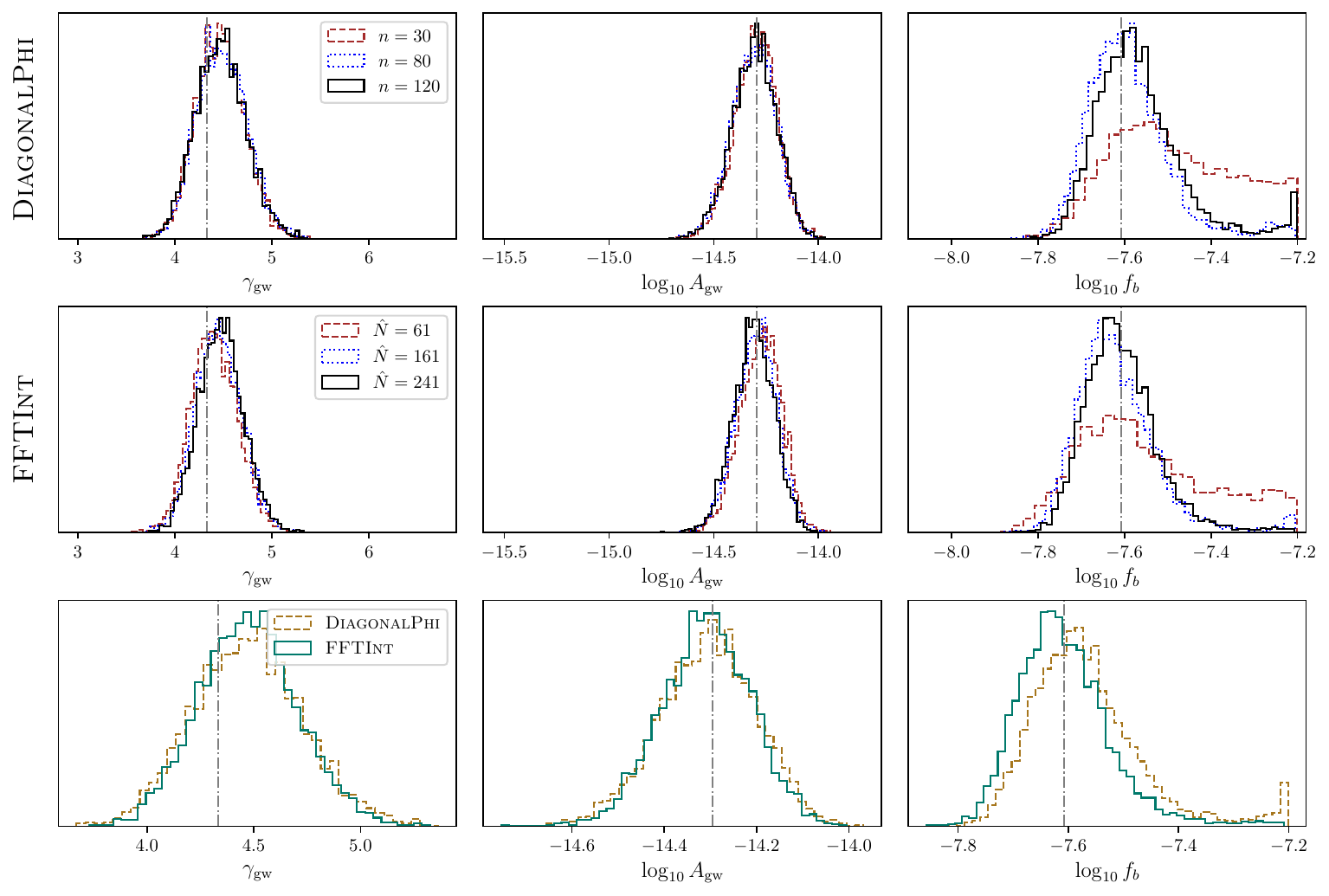}
    \end{center}
    \caption{Common red-noise posteriors 
    ($\gamma_\mathrm{gw}$, $\log_{10} A_\mathrm{gw}$, and $\log_{10} f_b$) for the full simulated NANOGrav dataset. The first row shows \DP\ posteriors with $n = 30, 80, 120$, the second row \FFT\ posteriors with $\hat{N} = 61, 161, 241$, and the third row compares the two methods for $\hat N = 241$ ($n=120$). Dashed--dotted lines show injected values.}
    \label{fig:corner}
\end{figure*}

In Fig.~\ref{fig:corner} we show $\gamma_\mathrm{gw}$, $\log_{10} A_\mathrm{gw}$, and $\log_{10} f_b$ posteriors, organized in a manner similar to Fig.~\ref{fig:pulsar}.
For both methods, the number of coarse-time nodes (frequency bins) has no discernible impact on the estimation of $\g_\mathrm{gw}$ and $A_\mathrm{gw}$. However, the posterior distribution of the break frequency $f_b$ becomes broader as the number of nodes decreases.
The third row of Fig.~\ref{fig:corner} shows no significant differences between the two methods for the common red-noise process.

Additionally, \FFT\ can be used for common processes with inter-pulsar correlations, such as the putative GWB studied in \cite{NANOGrav:2023gor}. We have verified that the same qualitative behavior occurs for that signal.

Because this analysis considers a single simulated realization, it is not an ensemble calibration study and does not establish statistical bias or coverage. Here we use the comparison only to identify method-dependent posterior shifts and agreement with the injected values in the realization considered.

\section{Conclusions}
\label{sec:conclusion}

GWB searches with PTAs rely critically on the accurate modeling of the Gaussian-process covariance of signal and noises alike.
In this paper we have shown evidence that the standard approximation (low-rank Fourier reconstruction with a diagonal Fourier-domain prior) overlooks important frequency correlations induced by the finite observation window, and can produce method-dependent shifts in parameter estimation.
We have described a more refined formulation of covariance, achieved through the application of the fast Fourier transform combined with a localized interpolation scheme, which allows us to efficiently approximate the true covariance matrix while avoiding Gibbs artifacts.

We demonstrated the limitations of the diagonal approximation by comparing it to our new method in controlled numerical tests with analytic covariance kernels (Sec.\ \ref{sec:comparison}) and through the analysis of a simulated dataset inspired by the NANOGrav 15-year data release (Sec.\ \ref{PTA}).
For smooth, low-frequency--dominated spectra the diagonal approximation is adequate, particularly after timing-model projection; for spectra with sharp in-band features it fails, with errors of order unity even after that projection (Sec.\ \ref{sec:stresstest}).
The diagonal approximation induces small errors in the estimation of the individual pulsars' timing-noise parameters for shallow-spectrum pulsars, but it does not produce discernible method-dependent shifts in the common-process parameters in this realization.

However, as datasets grow in size and sensitivity (by consolidating data from multiple PTAs and with next-generation facilities such as the SKA), the tolerance for mismodeling will shrink. In this context, accurate covariance modeling becomes essential. Our method offers a computationally tractable approach, enabling more precise and reliable inference without prohibitive computational costs, and it is already available with the data-analysis packages Enterprise \cite{enterprise} and Discovery \cite{discovery}.
We therefore advocate its adoption in future PTA investigations.

\begin{acknowledgments}
We are grateful to Katerina Chatziioannou, Aaron Johnson, Javier Roulet, Bruce Allen, Joe Romano, Ken Olum, and Heling Deng for many useful discussions.
M.C.\ is funded by the European Union under the Horizon Europe's Marie Sklodowska-Curie project~101065440.
\end{acknowledgments}

\appendix
\section{Conventions for an orthogonal reduced Fourier basis}
\label{app:ortho-fourier}

In Fig.~\ref{fig:signal-rec} we compare the reconstruction of a stochastic signal with the vector bases that underlie the \FFT\ and \SP\ covariance-matrix approximations. In that figure it can be seen that the two reconstructions intersect at all the nodes of the sparse time-domain basis. This would not be the case if we had used the usual conventions in the PTA literature, where the Fourier frequency spacing is set as $\Delta f = 1/T$. Furthermore, although we define the sparse time-domain basis for $t \in [2\,000, 6\,000]$, we can see that the periodicity of the signal in Figure~\ref{fig:signal-rec} is \emph{slightly} larger than $4\,000$. Those two subtleties are related, and arise from the fact that in the PTA literature the frequency spacing is set slightly differently from standard usage.

We start by defining the sparse grid on $t \in [t_{\text min}, t_{\text max}]$, where we denote the number of nodes as $\hat{N}$ as before. Then $t_{j} = t_{\text min} + j \Delta t$, with $j = 0, 1, \ldots, \hat{N}-1$. On such a grid, an orthogonal Fourier basis can be constructed on the frequencies
\be
f_{k} = k \Delta f, \quad \text{with} \quad
\Delta f = \frac{1}{\hat{N}\Delta t}.
\label{orthogonalfreqs}
\ee
This implies that the \emph{effective} length of the dataset is $T_\mathrm{eff}=\hat{N}\Delta t$, which is longer than the difference between the first and last observation: $t_{\text max} - t_{\text min} = (\hat{N}-1)\Delta t$. In the PTA literature, the frequency spacing is traditionally $\Delta f_\mathrm{trad} = 1/T = 1/(t_{\text max} - t_{\text min})$, which makes the Fourier basis non-orthogonal on regularly sampled data.

The \SP\ analyses in this manuscript are set up with the orthogonal frequency spacing of Eq.~\eqref{orthogonalfreqs}. This also implies that the effective periodicity is slightly larger than $t_{\text max} - t_{\text min}$. An additional advantage is that the model does not require the stochastic signal to have the same response at the first and last observation.

\bibliographystyle{utphys}
\bibliography{biblio}

\providecommand{\noopsort}[1]{}\providecommand{\singleletter}[1]{#1}%
\providecommand{\href}[2]{#2}\begingroup\raggedright\begin{thebibliography}{10}

\bibitem{romani1989timing}
R.~W. Romani, ``Timing a millisecond pulsar array,'' in {\em Timing Neutron
  Stars}, H.~{\"O}gelman and E.~P.~J. Heuvel, eds., pp.~113--117.
\newblock Springer, 1989.

\bibitem{fb90}
R.~S. {Foster} and D.~C. {Backer}, ``{Constructing a Pulsar Timing Array},''
  \href{http://dx.doi.org/10.1086/169195}{{\em \apj} {\bfseries 361} (Sept.,
  1990) 300}.

\bibitem{hd83}
R.~W. {Hellings} and G.~S. {Downs}, ``{Upper limits on the isotropic
  gravitational radiation background from pulsar timing analysis},''
  \href{http://dx.doi.org/10.1086/183954}{{\em Astrophys. J. Lett.} {\bfseries
  265} (Feb., 1983) L39--L42}.

\bibitem{rsg2015}
P.~A. Rosado, A.~Sesana, and J.~Gair, ``Expected properties of the first
  gravitational wave signal detected with pulsar timing arrays,''
  \href{http://dx.doi.org/10.1093/mnras/stv1098}{{\em Mon. Not. Roy. Astron.
  Soc.} {\bfseries 451} no.~3, (06, 2015) 2417--2433}.

\bibitem{2009PhRvD..79h4030A}
M.~{Anholm}, S.~{Ballmer}, J.~D.~E. {Creighton}, L.~R. {Price}, and
  X.~{Siemens}, ``{Optimal strategies for gravitational wave stochastic
  background searches in pulsar timing data},''
  \href{http://dx.doi.org/10.1103/PhysRevD.79.084030}{{\em \prd} {\bfseries 79}
  no.~8, (Apr., 2009) 084030}, \href{http://arxiv.org/abs/0809.0701}{{\ttfamily
  arXiv:0809.0701 [gr-qc]}}.

\bibitem{2015PhRvD..91d4048C}
S.~J. {Chamberlin}, J.~D.~E. {Creighton}, X.~{Siemens}, P.~{Demorest},
  J.~{Ellis}, L.~R. {Price}, and J.~D. {Romano}, ``{Time-domain implementation
  of the optimal cross-correlation statistic for stochastic gravitational-wave
  background searches in pulsar timing data},''
  \href{http://dx.doi.org/10.1103/PhysRevD.91.044048}{{\em \prd} {\bfseries 91}
  no.~4, (Feb., 2015) 044048}, \href{http://arxiv.org/abs/1410.8256}{{\ttfamily
  arXiv:1410.8256 [astro-ph.IM]}}.

\bibitem{2018PhRvD..98d4003V}
S.~J. {Vigeland}, K.~{Islo}, S.~R. {Taylor}, and J.~A. {Ellis},
  ``{Noise-marginalized optimal statistic: A robust hybrid frequentist-Bayesian
  statistic for the stochastic gravitational-wave background in pulsar timing
  arrays},'' \href{http://dx.doi.org/10.1103/PhysRevD.98.044003}{{\em \prd}
  {\bfseries 98} no.~4, (Aug., 2018) 044003},
  \href{http://arxiv.org/abs/1805.12188}{{\ttfamily arXiv:1805.12188
  [astro-ph.IM]}}.

\bibitem{Romano:2016dpx}
J.~D. Romano and N.~J. Cornish, ``{Detection methods for stochastic
  gravitational-wave backgrounds: a unified treatment},''
  \href{http://dx.doi.org/10.1007/s41114-017-0004-1}{{\em Living Rev. Rel.}
  {\bfseries 20} no.~1, (2017) 2},
  \href{http://arxiv.org/abs/1608.06889}{{\ttfamily arXiv:1608.06889 [gr-qc]}}.

\bibitem{taylor2021nanohertz}
S.~R. Taylor, {\em Nanohertz Gravitational Wave Astronomy}.
\newblock CRC Press, Boca Raton, FL, 2021.

\bibitem{vanHaasterenVallisneri2015}
R.~{van Haasteren} and M.~{Vallisneri}, ``{Low-rank approximations for large
  stationary covariance matrices, as used in the Bayesian and
  generalized-least-squares analysis of pulsar-timing data},''
  \href{http://dx.doi.org/10.1093/mnras/stu2157}{{\em Mon. Not. Roy. Astron.
  Soc.} {\bfseries 446} no.~2, (Jan., 2015) 1170--1174},
  \href{http://arxiv.org/abs/1407.6710}{{\ttfamily arXiv:1407.6710
  [astro-ph.IM]}}.

\bibitem{Hazboun:2019vhv}
J.~S. Hazboun, J.~D. Romano, and T.~L. Smith, ``{Realistic sensitivity curves
  for pulsar timing arrays},''
  \href{http://dx.doi.org/10.1103/PhysRevD.100.104028}{{\em Phys. Rev. D}
  {\bfseries 100} no.~10, (2019) 104028},
  \href{http://arxiv.org/abs/1907.04341}{{\ttfamily arXiv:1907.04341 [gr-qc]}}.

\bibitem{Depta:2024ykq}
P.~F. {Depta}, V.~{Domcke}, G.~{Franciolini}, and M.~{Pieroni}, ``{Pulsar
  timing array sensitivity to anisotropies in the gravitational wave
  background},'' \href{http://dx.doi.org/10.1103/PhysRevD.111.083039}{{\em
  \prd} {\bfseries 111} no.~8, (Apr., 2025) 083039},
  \href{http://arxiv.org/abs/2407.14460}{{\ttfamily arXiv:2407.14460
  [astro-ph.CO]}}.

\bibitem{Pitrou:2024scp}
C.~{Pitrou} and G.~{Cusin}, ``{Mitigating cosmic variance in the Hellings-Downs
  curve: A cosmic microwave background analogy},''
  \href{http://dx.doi.org/10.1103/PhysRevD.111.083544}{{\em \prd} {\bfseries
  111} no.~8, (Apr., 2025) 083544},
  \href{http://arxiv.org/abs/2412.12073}{{\ttfamily arXiv:2412.12073 [gr-qc]}}.

\bibitem{Bernardo:2024tde}
R.~C. {Bernardo} and K.-W. {Ng}, ``{Accurate pulsar timing array residual
  variances and correlation of the stochastic gravitational-wave background},''
  \href{http://dx.doi.org/10.1103/8cvw-fwx9}{{\em \prd} {\bfseries 113} no.~2,
  (Jan., 2026) L021501}, \href{http://arxiv.org/abs/2409.01218}{{\ttfamily
  arXiv:2409.01218 [astro-ph.CO]}}.

\bibitem{enterprise}
J.~Ellis, M.~Vallisneri, R.~van Haasteren, {\em et~al.}, ``Enterprise:
  v3.5.0,'' July, 2026.
\newblock \url{https://doi.org/10.5281/zenodo.21360719}.

\bibitem{discovery}
M.~Vallisneri, P.~M. Meyers, D.~Wright, A.~D. Johnson, J.~G. Baier, and R.~van
  Haasteren, ``{DISCOVERY}: the next-generation pulsar-timing-array
  data-analysis package,'' Nov., 2025.
\newblock \url{https://doi.org/10.5281/zenodo.17711454}.

\bibitem{jax2018github}
J.~Bradbury, R.~Frostig, P.~Hawkins, M.~J. Johnson, C.~Leary, D.~Maclaurin,
  G.~Necula, A.~Paszke, J.~Vander{P}las, S.~Wanderman-{M}ilne, and Q.~Zhang,
  ``{JAX}: composable transformations of {P}ython+{N}um{P}y programs,'' 2018.
\newblock \url{http://github.com/google/jax}.

\bibitem{baydin2018automatic}
A.~G. Baydin, B.~A. Pearlmutter, A.~A. Radul, and J.~M. Siskind, ``Automatic
  differentiation in machine learning: a survey,'' {\em Journal of Machine
  Learning Research} {\bfseries 18} no.~153, (2018) 1--43.

\bibitem{hoffman2014no}
M.~D. Hoffman and A.~Gelman, ``The {No-U-Turn} sampler: adaptively setting path
  lengths in {Hamiltonian} {Monte} {Carlo},'' {\em J. Mach. Learn. Res.}
  {\bfseries 15} no.~1, (2014) 1593--1623.

\bibitem{phan2019composable}
D.~Phan, N.~Pradhan, and M.~Jankowiak, ``Composable effects for flexible and
  accelerated probabilistic programming in {NumPyro},''
  \href{http://arxiv.org/abs/1912.11554}{{\ttfamily arXiv:1912.11554}}.

\bibitem{bingham2019pyro}
E.~Bingham, J.~P. Chen, M.~Jankowiak, F.~Obermeyer, N.~Pradhan, T.~Karaletsos,
  R.~Singh, P.~A. Szerlip, P.~Horsfall, and N.~D. Goodman, ``Pyro: Deep
  universal probabilistic programming,'' {\em J. Mach. Learn. Res.} {\bfseries
  20} (2019) 28:1--28:6. \url{http://jmlr.org/papers/v20/18-403.html}.

\bibitem{2009MNRAS.395.1005V}
R.~{van Haasteren}, Y.~{Levin}, P.~{McDonald}, and T.~{Lu}, ``{On measuring the
  gravitational-wave background using Pulsar Timing Arrays},''
  \href{http://dx.doi.org/10.1111/j.1365-2966.2009.14590.x}{{\em Mon. Not. Roy.
  Astron. Soc.} {\bfseries 395} no.~2, (May, 2009) 1005--1014},
  \href{http://arxiv.org/abs/0809.0791}{{\ttfamily arXiv:0809.0791
  [astro-ph]}}.

\bibitem{Allen:2024uqs}
B.~Allen and J.~D. Romano, ``{Optimal Reconstruction of the Hellings and Downs
  Correlation},'' \href{http://dx.doi.org/10.1103/PhysRevLett.134.031401}{{\em
  Phys. Rev. Lett.} {\bfseries 134} no.~3, (2025) 031401},
  \href{http://arxiv.org/abs/2407.10968}{{\ttfamily arXiv:2407.10968 [gr-qc]}}.

\bibitem{2023ApJ...951L...9A}
G.~{Agazie} {\em et~al.}, ``{The NANOGrav 15 yr Data Set: Observations and
  Timing of 68 Millisecond Pulsars},''
  \href{http://dx.doi.org/10.3847/2041-8213/acda9a}{{\em Astrophys. J. Lett.}
  {\bfseries 951} no.~1, (July, 2023) L9},
  \href{http://arxiv.org/abs/2306.16217}{{\ttfamily arXiv:2306.16217
  [astro-ph.HE]}}.

\bibitem{Lentati:2012xb}
L.~Lentati, P.~Alexander, M.~P. Hobson, S.~Taylor, J.~Gair, S.~T. Balan, and
  R.~van Haasteren, ``{Hyper-efficient model-independent Bayesian method for
  the analysis of pulsar timing data},''
  \href{http://dx.doi.org/10.1103/PhysRevD.87.104021}{{\em Phys. Rev. D}
  {\bfseries 87} no.~10, (2013) 104021},
  \href{http://arxiv.org/abs/1210.3578}{{\ttfamily arXiv:1210.3578
  [astro-ph.IM]}}.

\bibitem{vanHaasteren:2012hj}
R.~van Haasteren and Y.~Levin, ``{Understanding and analysing time-correlated
  stochastic signals in pulsar timing},''
  \href{http://dx.doi.org/10.1093/mnras/sts097}{{\em Mon. Not. Roy. Astron.
  Soc.} {\bfseries 428} (2013) 1147},
  \href{http://arxiv.org/abs/1202.5932}{{\ttfamily arXiv:1202.5932
  [astro-ph.IM]}}.

\bibitem{2014PhRvD..90j4012V}
R.~{van Haasteren} and M.~{Vallisneri}, ``{New advances in the Gaussian-process
  approach to pulsar-timing data analysis},''
  \href{http://dx.doi.org/10.1103/PhysRevD.90.104012}{{\em \prd} {\bfseries 90}
  no.~10, (Nov., 2014) 104012},
  \href{http://arxiv.org/abs/1407.1838}{{\ttfamily arXiv:1407.1838 [gr-qc]}}.

\bibitem{NANOGrav:2023gor}
G.~Agazie {\em et~al.}, ``{The NANOGrav 15 yr Data Set: Evidence for a
  Gravitational-wave Background},''
  \href{http://dx.doi.org/10.3847/2041-8213/acdac6}{{\em Astrophys. J. Lett.}
  {\bfseries 951} no.~1, (2023) L8},
  \href{http://arxiv.org/abs/2306.16213}{{\ttfamily arXiv:2306.16213
  [astro-ph.HE]}}.

\bibitem{EPTA:2023fyk}
J.~Antoniadis {\em et~al.}, ``{The second data release from the European Pulsar
  Timing Array. III. Search for gravitational wave signals},''
  \href{http://dx.doi.org/10.1051/0004-6361/202346844}{{\em Astron. Astrophys.}
  {\bfseries 678} (Oct., 2023) A50},
  \href{http://arxiv.org/abs/2306.16214}{{\ttfamily arXiv:2306.16214
  [astro-ph.HE]}}.

\bibitem{Reardon:2023gzh}
D.~J. Reardon {\em et~al.}, ``{Search for an Isotropic Gravitational-wave
  Background with the Parkes Pulsar Timing Array},''
  \href{http://dx.doi.org/10.3847/2041-8213/acdd02}{{\em Astrophys. J. Lett.}
  {\bfseries 951} no.~1, (2023) L6},
  \href{http://arxiv.org/abs/2306.16215}{{\ttfamily arXiv:2306.16215
  [astro-ph.HE]}}.

\bibitem{MilesShannon2025}
M.~T. {Miles} {\em et~al.}, ``{The MeerKAT Pulsar Timing Array: the first
  search for gravitational waves with the MeerKAT radio telescope},''
  \href{http://dx.doi.org/10.1093/mnras/stae2571}{{\em Mon. Not. Roy. Astron.
  Soc.} {\bfseries 536} no.~2, (Jan., 2025) 1489--1500},
  \href{http://arxiv.org/abs/2412.01153}{{\ttfamily arXiv:2412.01153
  [astro-ph.HE]}}.

\bibitem{Xu:2023wog}
H.~Xu {\em et~al.}, ``{Searching for the Nano-Hertz Stochastic Gravitational
  Wave Background with the Chinese Pulsar Timing Array Data Release I},''
  \href{http://dx.doi.org/10.1088/1674-4527/acdfa5}{{\em Res. Astron.
  Astrophys.} {\bfseries 23} no.~7, (2023) 075024},
  \href{http://arxiv.org/abs/2306.16216}{{\ttfamily arXiv:2306.16216
  [astro-ph.HE]}}.

\bibitem{gibbs1898Natur..59..200G}
J.~W. {Gibbs}, ``{Fourier's Series},''
  \href{http://dx.doi.org/10.1038/059200b0}{{\em \nat} {\bfseries 59} no.~1522,
  (Dec., 1898) 200}.

\bibitem{wilbraham1848certain}
H.~Wilbraham, ``On a certain periodic function,'' {\em The Cambridge and Dublin
  Mathematical Journal} {\bfseries 3} (1848) 198--201.

\bibitem{strang1986}
G.~Strang, {\em {Introduction to Applied Mathematics}}.
\newblock Wellesley-Cambridge Press, Wellesley, MA, 1986.

\bibitem{2024PhRvD.109j3012J}
A.~D. Johnson {\em et~al.}, ``{NANOGrav 15-year gravitational-wave background
  methods},'' \href{http://dx.doi.org/10.1103/PhysRevD.109.103012}{{\em \prd}
  {\bfseries 109} no.~10, (May, 2024) 103012},
  \href{http://arxiv.org/abs/2306.16223}{{\ttfamily arXiv:2306.16223
  [astro-ph.HE]}}.

\end{thebibliography}\endgroup

\end{document}